\documentclass[aps,prb,twocolumn,showpacs,reprint]{revtex4-1}

\usepackage{amssymb,color}
\usepackage{graphicx}
\usepackage{amsmath}
\usepackage{hyperref} 

\begin{document}

\title{Electron and hole Hong-Ou-Mandel interferometry}

\author{T.~Jonckheere}
\email[Email address: ]{thibaut.jonckheere@cpt.univ-mrs.fr}
\affiliation{Centre de Physique Th\'eorique,
Aix-Marseille Universit\'e,  CNRS UMR 7332, 
	Case 907 Luminy, 13288 Marseille cedex 9, France}
\author{J.~Rech}
\affiliation{Centre de Physique Th\'eorique, 
Aix-Marseille Universit\'e, CNRS UMR 7332, 
	Case 907 Luminy, 13288 Marseille cedex 9, France}
\author{C.~Wahl}
\affiliation{Centre de Physique Th\'eorique,  
Aix-Marseille Universit\'e, CNRS UMR 7332,
	Case 907 Luminy, 13288 Marseille cedex 9, France}
\author{T.~Martin}
\affiliation{Centre de Physique Th\'eorique,
Aix-Marseille Universit\'e,  CNRS UMR 7332, 
	Case 907 Luminy, 13288 Marseille cedex 9, France}

\date{\today}

\begin{abstract}
We consider the electronic analog of the quantum optics Hong-Ou-Mandel interferometer, in a realistic condensed matter device
based on single electron emission in chiral edge states. 
For electron-electron collisions, we show that the measurement of the zero-frequency current correlations at the output of a quantum point contact produces a dip giving precious information on the electronic wavepackets and coherence.
As a feature truly unique to Fermi statistics and condensed matter, we show that two-particle interferences 
between electron and hole in the Fermi sea can produce a positive peak in the current correlations, 
which we study for realistic experimental parameters.
\end{abstract}

\pacs{
	73.23.-b, 
	72.70.+m, 
	42.50.-p  
}

\maketitle

\section{Introduction}
The Hong-Ou-Mandel (HOM) interferometer is a celebrated tool of quantum optics, where 
two photons are sent on the two input channels of a beam splitter, 
while measuring the coincidence rate at the two outputs~\cite{grynberg_introduction_2010}.
A dip is observed when the photons are identical and arrive simultaneously at the splitter,
as they necessarily go to the same output because of bosonic statistics. 
Measuring this dip can give access to the time difference between the photons, and to the length
of the photon wavepacket~\cite{hong_measurement_1987}. Since the first HOM experiments,
many works have used this interferometer, e.g. to characterize single photon sources and photon
indistinguishability~\cite{patel_quantum_2010,jha_temporal_2008,beugnon_quantum_2006},
 to demonstrate the control of interferences with a resonant cavity~\cite{sagioro_time_2004,zavatta_recurrent_2004},
to perform interference between two photon pairs~\cite{cosme_hong-ou-mandel_2008}, etc.

Single electron sources in condensed matter physics are now available: electrons can be emitted 
periodically into edge states 
of the quantum Hall effect (QHE)~\cite{feve_-demand_2007,mahe_current_2010,leicht_generation_2011}.
Such sources open the way to performing analogs of quantum optics experiments, such as Hanbury-Brown Twiss (HBT) interferometry~\cite{martin_wave-packet_1992,bocquillon_electron_2012}.
 As electrons obey the fermionic statistics, and are subject to Coulomb interactions, 
one expects to observe fundamental
departures from photon measurements. In particular, electron vacancies
in the Fermi sea (the set of low energy states filled with electrons) create ``holes'', 
which have a charge opposite to that of the electron. In two-fermion 
interference experiments, it is essential to understand the 
role of the Fermi sea, of electron/hole pair creation and propagation.

The electronic analog of the HOM experiment in condensed matter
has so far eluded a complete theoretical description~\cite{burkard_lower_2003,giovannetti_electronic_2006,
feve_quantum_2008,olkhovskaya_shot_2008,moskalets_spectroscopy_2011}.
Contrary to photons, two identical electrons arriving simultaneously
at the splitter exit in two opposite channels because of Fermi statistics. 
Surprisingly, this leads to a signature similar to that of photons, albeit on a different physical quantity.
For electrons, it is the current-current cross-correlation at zero frequency which exhibits a dip (in absolute value, as cross correlations are negative) when the time difference between the electrons is varied. 
Using analytical calculations describing a realistic model of the emitter,
we show that two-electron interferences bear strong similarities with that of photons.
The shape of the dip provides valuable information on the wavepackets.
We next consider the case of interferences between an electron and a hole, which has no 
counterpart with photons: positive interferences (a peak rather than a dip) are obtained,
which depend on the energy overlap between the electron and hole wavepackets, and which is strongly affected by
temperature. 

This article is organized as follows. In section~\ref{sec:setup}, we describe the setup and the formalism of our calculations.
Section~\ref{sec:eecol} is devoted to the results we obtain for electron-electron collisions. In section~\ref{sec:ehcol},
we consider the case of a collision between a hole and an electron, and conclusions are given in section~\ref{sec:conclusion}.

\section{Setup and formalism}
\label{sec:setup}
\begin{figure}
\includegraphics[width=7.cm]{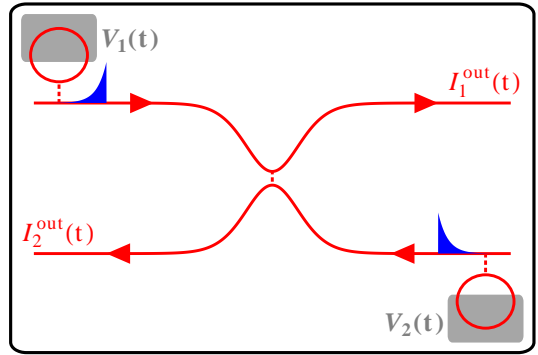}
\caption{(Color online) The device. Two chiral edge states meet at a quantum point contact. In each
edge state, a single electron source injects an electron when a time-dependent voltage
$V(t)$ is applied at the source. Cross-correlations at the two outputs are measured
as a function of the time difference between the electrons which are emitted by each source.
 }
\label{fig:device}
\end{figure}
In Fig.~\ref{fig:device}, two counter-propagating edge states meet at a quantum
point contact (QPC).~\cite{olkhovskaya_shot_2008} A single electron emitter
is connected to each incoming edge state, and single electrons can be injected with a controlled 
time difference in the two edge states. The current correlations
are measured at the two output of the QPC, as a function of the time difference between the two injections.
In existing experiments, the emission of a single particle is 
repeated periodically~\cite{feve_-demand_2007,mahe_current_2010,parmentier_current_2011}.

The zero-frequency current correlations between the outputs read:
\begin{multline}
S_{12}^{out} = \int dt \, dt'  \Big(\left \langle I_1^{out}(t) I_2^{out}(t') \right \rangle \\
          - \left \langle I_1^{out}(t) \right \rangle \left \langle I_2^{out}(t') \right \rangle   \Big) \;\;,  
\label{eq:noisestart}  
\end{multline}
where $I_1^{out}(t)$ and $I_2^{out}(t)$ are the currents outgoing from the QPC. 
With a linear dispersion, the currents depend on $x-t$ only 
(we set the Fermi velocity $v_F=1$), and
we compute the current at the immediate output of the QPC ($x=0^+$, where for convenience, we use for each edge
a x-axis pointing in the propagation direction).
The QPC is described by its scattering matrix which expresses the outgoing fields in terms
of the incoming ones:
\begin{equation}
\begin{pmatrix}  \psi_{2}^{out} (t) \\ \psi_{1}^{out} (t) \end{pmatrix} 
 = 
 \begin{pmatrix} \sqrt{\mathcal T} & i \sqrt{\mathcal R} \\ i \sqrt{\mathcal R} &  \sqrt{\mathcal T} \end{pmatrix}
 \begin{pmatrix}  \psi_{2}^{in} (t) \\ \psi_{1}^{in} (t) \end{pmatrix} \; ,
\end{equation}
where ${\mathcal T}$ and ${\mathcal R}=1-{\mathcal T}$ are the transmission and reflection probabilities.
 Expressing the currents operators
at the output of the QPC in terms of the incoming ones, and dropping the $in$ superscript gives
\begin{align}
I_{1}^{out}(t) &=  {\mathcal T} I_1(t) +  {\mathcal R} I_2(t) + i e 
           \sqrt{\mathcal{R} \mathcal{T}}(\psi^{\dagger}_1 \psi_2 - \psi^{\dagger}_2 \psi_1)(t)  \nonumber \\ 
I_{2}^{out}(t) &=  {\mathcal R} I_1(t) + {\mathcal T} I_2(t) - i e 
           \sqrt{\mathcal{R} \mathcal{T}} (\psi^{\dagger}_1 \psi_2 - \psi^{\dagger}_2 \psi_1)(t)  \;.    \nonumber    
\end{align} 
Using these expressions in Eq.~(\ref{eq:noisestart})
allows us to express the noise in terms of the incoming operators only~\cite{grenier_single-electron_2011}
\begin{equation}
S_{12}^{out} =  \mathcal{R T} \left(S_{11} + S_{22} - \mathcal{M} \right) \;,
\label{eq:S12}
\end{equation}
where the first two terms are the autocorrelation noise on the two incoming edges.
The last term combines averages on both incoming edges:
\begin{align}
\mathcal{M} = e^2 \int dt dt' \;  &\Big[
    \langle \psi_1(t) \psi_1^{\dagger}(t') \rangle \langle \psi_2^{\dagger}(t) \psi_2(t') \rangle \nonumber \\
    & +  \langle \psi_1^{\dagger}(t) \psi_1(t') \rangle \langle \psi_2(t) \psi_2^{\dagger}(t') \rangle \Big]\; .
\end{align}
These averages are performed on the state $|\Psi\rangle$ of the edges,
and correspond to electronic coherence functions which generalize the optical coherences~\cite{grenier_single-electron_2011}.
The transmission $\mathcal{T}$  simply
acts as a trivial prefactor for the current correlations in $S_{12}^{out}$. 

\section{Electron-electron collisions}
\label{sec:eecol}
In order to get analytical formulas for the HOM dip, we perform
calculations where a single electron, with a given wavepacket, is added to each edge.
These analytical formulas are next compared with Floquet calculations for a periodic source,
where no simple analytical formulas are available in our case (note that anayltical formulas can be obtained
in the case of a slow sinuoidal drive, see Ref.~\onlinecite{olkhovskaya_shot_2008}). 
 
 When one electron with wavefunction $\phi_{1,2}(x)$ is added to each edge, the states describing the edges are:
\begin{equation}
| \Psi_{1,2} \rangle = \int dx \; \phi_{1,2}(x) \; \psi^{\dagger}_{1,2}(x) \; | 0 \rangle
\label{eq:Psie}
\end{equation}
where $| 0 \rangle$ stands for the Fermi sea at temperature $T$ of the edge. In the case of identical wavepackets 
$\phi_{1}(x)\!\!=\!\!\phi_{2}(x)\!\!=\!\!\phi(x)$, reaching
the QPC with a time difference $\delta t$, we get for the noise:
\begin{equation}
\frac{S_{12}^{out}(\delta t)}{2 \mathcal{S}_{HBT}} =
1 - \left| \frac{\int_0^\infty dk
  |\phi(k)|^2 e^{-i k \delta t} (1-f_k)^2}
  {\int_0^\infty dk  |\phi(k)|^2 (1-f_k)^2} \right|^2 
\label{eq:HOMee1}
\end{equation}
where $f_k = 1/(1+e^{(k-k_F)/T})$ is the Fermi distribution,
$\phi(k)$ is the normalized wavefunction in momentum space. $ \mathcal{S}_{HBT}$ is the noise in the 
HBT configuration, where only a single electron is emitted towards the QPC~\cite{bocquillon_electron_2012} :
\begin{equation}
 \mathcal{S}_{HBT}   =
-e^2 \mathcal{R} \mathcal{T}
\left(
\frac{\int_0^\infty dk
  |\phi(k)|^2  (1-f_k)^2}
  {\int_0^\infty dk  |\phi(k)|^2 (1-f_k)} \right)^2  \; .
\label{eq:SHBT}
\end{equation}
Eq.~(\ref{eq:HOMee1}) shows immediately that when the two identical electrons reach the QPC
simultaneously ($\delta t=0$), the noise is zero, as expected from Fermi statistics. On the other hand, for large values
of $\delta t$  (much larger than the inverse of the typical width of $\phi(k)$), the numerator in Eq.~(\ref{eq:HOMee1}) vanishes,
 and the noise is given by $ 2 \mathcal{S}_{HBT}$, the sum of the noise of the two electrons
taken independently. At low temperature, when the wavepacket $\phi(k)$ has weight  
above the Fermi level only, the noise can be simplified further:
\begin{equation}
\frac{S_{12}^{out}(\delta t)}{2 \mathcal{S}_{HBT}} =
  1- \left| \int dx \; \phi(x) \, \phi^*(x+\delta t)\right|^2 \;.
  \label{eq:HOMeerealspace}
\end{equation}
This expression is reminiscent of the one obtained in optics, where the shape of the HOM dip is given by the
self-convolution of the photon wavepacket~\cite{hong_measurement_1987}.

Eqs.~(\ref{eq:HOMee1}) and (\ref{eq:HOMeerealspace}) are easily generalized to the case of two different wavepackets.
Specific formulas of the HOM dip can be obtained analytically for different shapes of wavepackets.
We concentrate here on Lorentzian wavepackets in energy space
\begin{equation}
\phi_{\Gamma}(k) = \frac{\sqrt{\Gamma}}{\sqrt{\pi}} \frac{1}{(k-k_0)+i \Gamma}
\end{equation}
 which 
corresponds to the emission by the discrete level of a quantum dot at energy $k_0$, as found in the single-electron
emitter. The real space profile of this wavepacket is exponential (see Fig~\ref{fig:device}):
\begin{equation}
\phi_{\Gamma}(x) = \sqrt{2 \Gamma} e^{i k_0 x} e^{\Gamma x} \theta(-x) 
\label{eq:Expwavepacket}
\end{equation}
where $\theta(x)$ is the Heavyside function.
The noise at zero temperature for the case of two Lorentzian wavepackets, 
centered at the same $k_0$ but with different widths $\Gamma_{1,2}$
reads:
\begin{equation}
\frac{S_{12}^{out}(\delta t)}{2 \mathcal{S}_{HBT}} = 1 -
 \frac{4 \Gamma_1 \Gamma_2}{(\Gamma_1 + \Gamma_2)^2}  
  \Big[ \theta(\delta t) e^{-2 \Gamma_1 \delta t} + \theta(-\delta t) e^{2 \Gamma_2 \delta t} \Big] ,
\label{eq:HOMeeExp}
\end{equation}
when wavepacket $2$ reaches the QPC at a time $\delta t$ after wavepacket $1$.
A remarkable feature of this HOM dip is its asymmetry: it has an exponential behavior, with different
time constants depending on the sign of $\delta t$. The contrast
of the HOM dip,  $4 \Gamma_1 \Gamma_2/(\Gamma_1 + \Gamma_2)^2$, is smaller than $1$,
 which reflects the fact that the two electrons are not identical. The same contrast was obtained
 in Ref.~\onlinecite{olkhovskaya_shot_2008} for a periodic source with a slow sinusoidal drive.
It is only in the case of identical wavepackets that a maximum contrast is recovered, 
and the HOM dip has a simple exponential form $1-e^{-2 \Gamma |\delta t|}$.
For arbitrary wavepackets, Eq.~(\ref{eq:HOMeerealspace}) shows that the asymmetry of the HOM dip is possible 
only if the wavepackets in real space have no mirror symmetry. This is clearly the case for exponential wavepackets,
but it is not true for instance for Lorentzians in real space. 
The asymmetry in the HOM dip thus provides information on the spatial symmetry of the 
electronic wavepackets. 
 
We now compare Eq.~(\ref{eq:HOMeeExp}) with the results of a Floquet calculation including 
the emission process from the single-electron emitters.
This emitter is based on the mesoscopic capacitor~\cite{feve_-demand_2007,mahe_current_2010,bocquillon_electron_2012}.
A complete description of this system, and of the Floquet scattering theory is available~\cite{parmentier_current_2011,moskalets_quantized_2008}. A quantum dot is connected through a QPC to the main edge state,
 and is capacitively coupled to a gate. The dot has discrete energy levels, whose width increases
 with the transparency of the QPC and with temperature. A time periodic voltage $V(t)$ on the gate is used
 to create an oscillation of the dot levels. First the highest occupied level is put far above the Fermi
 energy, which causes the emission of one electron from the dot to the main edge; second this (now empty) level
 is put far below the Fermi level, which causes the emission of a hole in the main edge. 
 Several profiles of $V(t)$ are possible. The one which yields optimal emission 
 uses voltage steps, with an amplitude equal to the level spacing of the dot $\Delta$. 
 The electron is expected to be emitted with a Lorentzian energy profile, which reflects 
 the Lorentzian density of state of the dot level.
  \begin{figure}[!]
\includegraphics[width=9.cm]{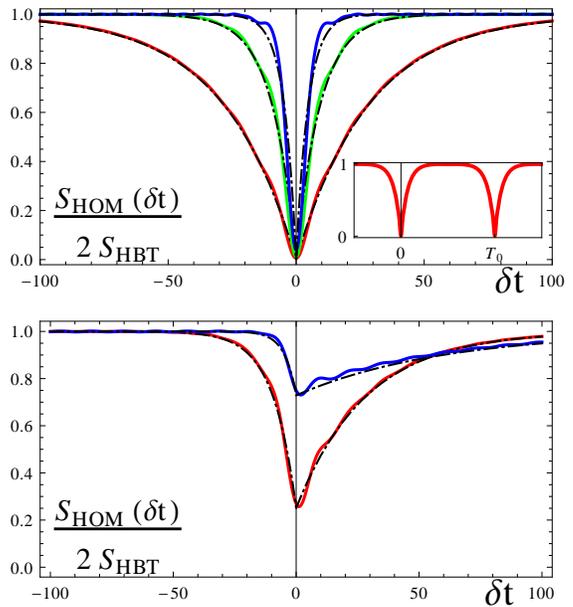}
\caption{(Color online) HOM dips as a function of the time difference $\delta t$, obtained from Floquet scattering matrix formalism in the optimal emission regime (full) and analytical predictions of Eq.~(\ref{eq:HOMeeExp}) for exponential wavepackets (dot-dashed). 
Upper panel: symmetric case, with emitter transparencies $D=0.2$, $0.5$ and $0.8$ (from outer to inner curve).
Inset: dips for $D=0.2$ on two periods of the applied voltage.
 Lower panel: asymmetric case, with transparencies
 $D_1=0.2$, $D_2=0.5$ (bottom curve) and $D_1=0.1$, $D_2=0.8$ (top curve).
 }
\label{fig:HOMexp}
\end{figure}
 
The dot is described by its scattering matrix, and the Floquet formalism is used to take advantage 
of the time periodicity~\cite{moskalets_floquet_2002} in order to compute the relevant 
quantities~\cite{parmentier_current_2011}.  
Fig.~\ref{fig:HOMexp} compares the results for the HOM dip with
the analytical formula of Eq.~(\ref{eq:HOMeeExp}). 
The period of the applied voltage $V(t)$ of the emitters is $T_0 = 400$ (in units of $\hbar/\Delta$), and temperature is chosen to 
be small ($T=0.01 \Delta$).
When $V(t)$ consists of sharp periodic steps (optimal emission) the
 emitting level is put alternatively at energy $+\Delta /2$ and $-\Delta /2$ with respect to the Fermi level.
 The upper panel shows the symmetric case, where the two emitters have identical parameters.
 For three different transparencies, we observe three different HOM dips
 with a maximum contrast (the minimum value is $0$ at $\delta t=0$). 
 The dip is broader for lower emitter transparency, which signals a broader wavepacket, as the electron takes
 a longer time to exit the dot.
 For the emitter
 operated in the optimal regime, it is known that the electron emission time as a function of the
 transparency $D$ is given by $\tau = (2\pi/\Delta) (1/D- 1/2)$~\cite{parmentier_current_2011}.
 We have used this value of the emission time (with $2 \Gamma=\tau^{-1}$) to plot the analytical
 predictions from Eq.~(\ref{eq:HOMeeExp}) (dot-dashed curves).
 The agreement is excellent, without any fitting parameters, especially in the low transparency regime where
  it is known that true single electron emission is achieved~\cite{mahe_current_2010, grenier_single-electron_2011}.

In the asymmetric case (lower panel), as
 the emitted electrons are not identical, the contrast is smaller than 1, and the HOM dips are
 asymmetric. The agreement with the analytical prediction is again very good, both for the 
 asymmetric shapes and for the value of the contrast.
    
\section{Electron-hole collisions} 
\label{sec:ehcol}
The results presented so far are quite similar to those obtained with photons in optics. However, the existence of the Fermi sea allows to create holes in it, which have no 
counterpart for photons. We study two-particle interferences of electrons and holes, injecting one 
electron in one branch, with a state given by Eq.~(\ref{eq:Psie}),
and one hole injected on the other branch, with a state
$| \Psi_{h} \rangle = \int dx \; \phi_{h}(x) \; \psi(x) \; | 0 \rangle$.
We get for the noise:
\begin{widetext}
\begin{equation*}
S_{12}^{out}(\delta t) = -e^2 \mathcal{R} \mathcal{T} \Bigg[
\left(
\frac{\int_0^\infty dk
  |\phi_e(k)|^2  (1-f_k)^2}
  {\int_0^\infty dk  |\phi_e(k)|^2 (1-f_k)} \right)^2 +
\left(
\frac{\int_0^\infty dk
  |\phi_h(k)|^2  f_{k}^2}
  {\int_0^\infty dk |\phi_h(k)|^2 f_{k}} \right)^2 
+
 2 \frac{ \left|\int_0^\infty \! dk \,
  \phi_{e}(k) \phi_{h}^*(k) e^{-i k \delta t} f_k (1-f_k)\right|^2}
  {\int_0^\infty \! dk  \, |\phi_e(k)|^2 (1-f_k) \int_0^\infty dk' |\phi_h(k')|^2 f_{k'}}
\Bigg]   
\end{equation*}
\end{widetext}
where the first two terms are the HBT noise of the injected electron and hole respectively, and the third term
is due to the electron-hole interferences. 
For the physics to be the most transparent,
we consider in the following the electron-hole symmetic case:
$\phi_{e}(k_F+\delta k) = \phi_{h}(k_F- \delta k)$.
In this case, the first two terms are equal to $\mathcal{S}_{HBT}$ (see Eq.(\ref{eq:SHBT})).
The expression of the noise then simplifies to a form similar to the electron-electron case: 
\begin{equation}
\frac{S_{12}^{out}(\delta t)}{2 \mathcal{S}_{HBT}} =
1 +  \left|\frac{ \int_0^\infty \! dk \,
  \phi_{e}(k) \phi_{h}^*(k) e^{-i k \delta t} f_k (1-f_k)}
  {\int_0^\infty \! dk  \, |\phi_e(k)|^2 (1-f_k)^2} \right|^2 \; .
\label{eq:HOMeh1}
\end{equation}
 Comparing
Eq.~(\ref{eq:HOMeh1}) with Eq.~(\ref{eq:HOMee1}), we notice important changes. 
First, the interferences contribute now with a positive sign to the noise, i.e. the opposite
of the electron-electron case. Electron-hole interferences produce a ``HOM peak'' rather
than a dip. Second, the value of this peak depends on the overlap of the electron and the hole 
wavepackets ($\phi_{e}(k) \phi_{h}^*(k)$), times the Fermi product $f_k (1-f_k)$.
This peak thus vanishes as $T\to 0$ since it requires a significant overlap between electron and hole wavepackets, a situation which only happens in an energy range $\sim k_B T$ around $k_F$, where electronic states are neither fully occupied nor empty.

 Note that the many-body state $|\Psi_e\rangle$ (or $|\Psi_h\rangle$)
 created by the application of the electron creation (or annihilation) operator 
as in Eq.~(\ref{eq:Psie}) is quite complex when the wavepacket $\phi_e(x)$ (or $\phi_h(x)$) has an important weight
close to the Fermi energy. Indeed, due to the changes imposed on the Fermi sea, 
many electron-hole pairs are created, and the state is not simply one electron (or one
hole) plus the unperturbed Fermi sea. 
The appearance of a positive HOM peak can be attributed to interferences between these
electron-hole pairs coming from the two branches of the setup.
  It is quite remarkable that eventually, the peak can simply be computed from
the overlap of the electron and hole wavepackets (see Eq.(\ref{eq:HOMeh1})).

We now consider the observation of these electron-hole interferences with realistic electron emitters using Floquet scattering theory.
The simultaneous arrival of an electron and a hole at the QPC happens quite naturally in this device. Indeed,
each emitter emits periodically an electron and then, half a period $T_0$ later, a ``hole''. 
If one applies a time-shift $T_0/2$ between the two emitters, one expects that an electron from 
one emitter, and a hole from the other one, will interfere at the QPC. The optimal emission regime of
Fig.~\ref{fig:HOMexp} does not show any sign of electron-hole interference, as can be seen in the inset 
of this figure (no peak at $t=T_0/2$). This is easily understood from Eq.~(\ref{eq:HOMeh1}), as
the electron (hole) wavepacket is a Lorentzian peaked at energy $\Delta/2$ ($-\Delta/2$), and the overlap of
the two wavepackets is negligible, even in the totally transparent case.

\begin{figure}
\includegraphics[width=5.cm]{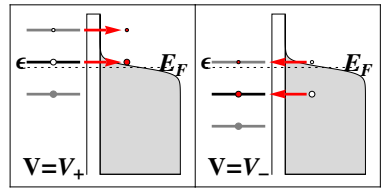}
\includegraphics[width=7.6cm]{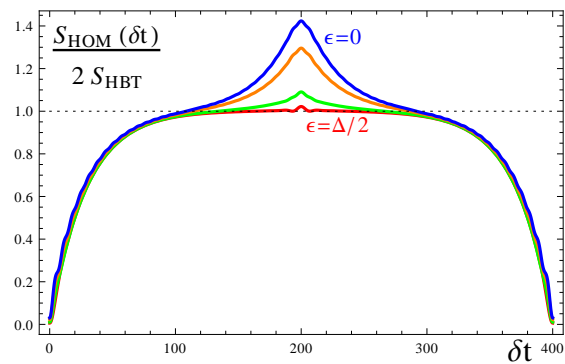}
\caption{(Color online) Upper panel:
electron (left) and hole (right) emission process, for a square voltage drive,
for the two positions of the dot levels (values $V_{+}$ and $V_{-}$ of the drive). The 
 position of the dot levels is parametrized by $\epsilon$ with respect to $E_F$.
 Bottom panel:
HOM peak for electron-hole collision, for a square voltage drive, at different level positions
$\epsilon=0.5$, 0.4, 0.25 and 0 (in units of $\Delta$) from the smallest to the largest peak.
$T=0.1 \Delta$ in both panels.
}
\label{fig:HOMElTrSq}
\end{figure}

We have considered two regimes where there is a significant overlap between electron and hole 
wavepackets, leading to observable electron-hole peak. In both cases, we have used a small but non-zero temperature
$T = 0.1 \Delta$, which is compatible with existing experimental values (typically $\Delta \simeq 2$K, and
$T \simeq 150$ mK~\cite{bocquillon_electron_2012}).
The first regime is the ``resonant emission'' regime, where the applied voltage still
consists of sharp steps with amplitude $\Delta$, but where the dot levels are shifted with respect to the optimal emission
regime (see the upper panel of Fig.~\ref{fig:HOMElTrSq}), such that there is always one dot level which is resonant with the Fermi level of the edge (the dot levels
go back and forth between energies $E_F + n \Delta$ and $E_F + (n+1) \Delta$, with $n$ an integer). 
Electron and hole emission thus happens for an important part at the Fermi energy, which leads to a significant
overlap between electron and hole (see Eq.(\ref{eq:HOMeh1})), and to the creation of electron-hole pairs~\cite{keeling_coherent_2008}.
 The results for the HOM noise are shown
in the bottom panel Fig.~\ref{fig:HOMElTrSq}, which have been computed for an emitter with transparency $D=0.2$. 
The different curves span the different cases between optimal emission (negligible peak at $t=T_0/2$)
and resonant emission (largest peak $t=T_0/2$).
 This peak has shape similar to the HOM dip, and it is due to the exponential profile of the
wavepackets in real space.

\begin{figure}
\includegraphics[width=7.6cm]{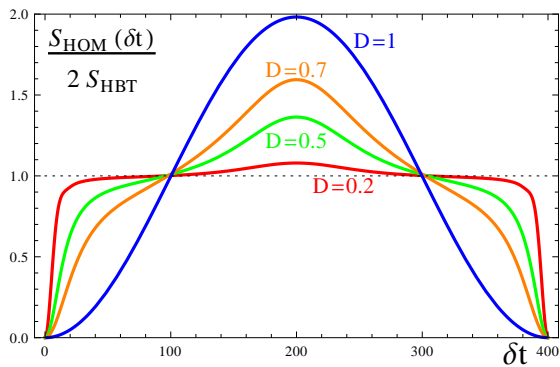}
\caption{(Color online) 
HOM peak for electron-hole collision, for a sine voltage drive, at different transparencies, 
$D=0.2$, 0.5, 0.7 and 1.0 (smaller to larger peak), with $T=0.1 \Delta$.}
\label{fig:HOMElTrSin}
\end{figure}

In the second regime the applied voltage has a sinusoidal 
dependence ($V(t) = (\Delta/2) \sin(2\pi t/T_0)$), rather than steps. 
The main emitting dot level now has a time dependence of the form $\sim E_F +  V(t)$,
which also creates energy distributions for the emitted electron and hole with substantial weight close to the 
Fermi energy. This weight increases with the dot transparency~\cite{bocquillon_electron_2012},
 and one can thus expect that the HOM peak
grows larger for transparencies closer to $1$. The result for the HOM noise in this case is shown in
Fig.~\ref{fig:HOMElTrSin} for different values of $D$.
We observe a small positive peak at $t=T_0/2=200$ for the smallest transparency, which increases
as $D$ is increased. For a totally transparent dot ($D=1$), the peak becomes as large as the
HOM dip (for interfences processes between two sinusoidal sources, see also Ref.~\onlinecite{rychkov_photon-assisted_2005}).


\section{Conclusions}
\label{sec:conclusion}
In conclusion, the fermionic HOM interferometer exhibits a dip in the zero-frequency current correlations when two electrons collide. Through analytic calculations, we were able to relate the shape of the dip to the properties of the electron wavepackets colliding. This showed good agreement with the more realistic treatment of the single-electron emitter via Floquet scattering theory, which also allows to consider more experimentally relevant situations.
We showed that interferences between an electron and a hole in the Fermi sea,
both close to the Fermi energy, can produce an HOM peak, which should be observable in future experimental setups involving two emitters. Future work should address the role of Coulomb interactions
 within the edge or between the two edges at the location of the QPC.
  
\acknowledgments  
This work was supported by the ANR grant ``1shot'' (ANR-2010-BLANC-0412).
We thank P. Degiovanni and G. F\`eve for valuable comments on the manuscript, 
and gratefully acknowledge useful discussions with the participants of the 1shot project.
 
\bibliography{../../mybib1}{}

\begin{thebibliography}{24}
\expandafter\ifx\csname natexlab\endcsname\relax\def\natexlab#1{#1}\fi
\expandafter\ifx\csname bibnamefont\endcsname\relax
  \def\bibnamefont#1{#1}\fi
\expandafter\ifx\csname bibfnamefont\endcsname\relax
  \def\bibfnamefont#1{#1}\fi
\expandafter\ifx\csname citenamefont\endcsname\relax
  \def\citenamefont#1{#1}\fi
\expandafter\ifx\csname url\endcsname\relax
  \def\url#1{\texttt{#1}}\fi
\expandafter\ifx\csname urlprefix\endcsname\relax\def\urlprefix{URL }\fi
\providecommand{\bibinfo}[2]{#2}
\providecommand{\eprint}[2][]{\url{#2}}

\bibitem[{\citenamefont{Grynberg et~al.}(2010)\citenamefont{Grynberg, Aspect,
  and Fabre}}]{grynberg_introduction_2010}
\bibinfo{author}{\bibfnamefont{G.}~\bibnamefont{Grynberg}},
  \bibinfo{author}{\bibfnamefont{A.}~\bibnamefont{Aspect}}, \bibnamefont{and}
  \bibinfo{author}{\bibfnamefont{C.}~\bibnamefont{Fabre}},
  \emph{\bibinfo{title}{Introduction to Quantum Optics: From the Semi-classical
  Approach to Quantized Light}} (\bibinfo{publisher}{Cambridge University
  Press}, \bibinfo{year}{2010}), ISBN \bibinfo{isbn}{0521551129}.

\bibitem[{\citenamefont{Hong et~al.}(1987)\citenamefont{Hong, Ou, and
  Mandel}}]{hong_measurement_1987}
\bibinfo{author}{\bibfnamefont{C.~K.} \bibnamefont{Hong}},
  \bibinfo{author}{\bibfnamefont{Z.~Y.} \bibnamefont{Ou}}, \bibnamefont{and}
  \bibinfo{author}{\bibfnamefont{L.}~\bibnamefont{Mandel}},
  \bibinfo{journal}{Phys. Rev. Lett.} \textbf{\bibinfo{volume}{59}},
  \bibinfo{pages}{2044} (\bibinfo{year}{1987}).

\bibitem[{\citenamefont{Patel et~al.}(2010)\citenamefont{Patel, Bennett,
  Cooper, Atkinson, Nicoll, Ritchie, and Shields}}]{patel_quantum_2010}
\bibinfo{author}{\bibfnamefont{R.~B.} \bibnamefont{Patel}},
  \bibinfo{author}{\bibfnamefont{A.~J.} \bibnamefont{Bennett}},
  \bibinfo{author}{\bibfnamefont{K.}~\bibnamefont{Cooper}},
  \bibinfo{author}{\bibfnamefont{P.}~\bibnamefont{Atkinson}},
  \bibinfo{author}{\bibfnamefont{C.~A.} \bibnamefont{Nicoll}},
  \bibinfo{author}{\bibfnamefont{D.~A.} \bibnamefont{Ritchie}},
  \bibnamefont{and} \bibinfo{author}{\bibfnamefont{A.~J.}
  \bibnamefont{Shields}}, \bibinfo{journal}{Nanotechnology}
  \textbf{\bibinfo{volume}{21}}, \bibinfo{pages}{274011}
  (\bibinfo{year}{2010}), ISSN \bibinfo{issn}{0957-4484, 1361-6528}.

\bibitem[{\citenamefont{Jha et~al.}(2008)\citenamefont{Jha,
  {O{\textquoteright}Sullivan}, Chan, and Boyd}}]{jha_temporal_2008}
\bibinfo{author}{\bibfnamefont{A.~K.} \bibnamefont{Jha}},
  \bibinfo{author}{\bibfnamefont{M.~N.}
  \bibnamefont{{O{\textquoteright}Sullivan}}},
  \bibinfo{author}{\bibfnamefont{K.~W.~C.} \bibnamefont{Chan}},
  \bibnamefont{and} \bibinfo{author}{\bibfnamefont{R.~W.} \bibnamefont{Boyd}},
  \bibinfo{journal}{Phys. Rev. A} \textbf{\bibinfo{volume}{77}},
  \bibinfo{pages}{021801} (\bibinfo{year}{2008}).

\bibitem[{\citenamefont{Beugnon et~al.}(2006)\citenamefont{Beugnon, Jones,
  Dingjan, Darqui\'{e}, Messin, Browaeys, and Grangier}}]{beugnon_quantum_2006}
\bibinfo{author}{\bibfnamefont{J.}~\bibnamefont{Beugnon}},
  \bibinfo{author}{\bibfnamefont{M.~P.~A.} \bibnamefont{Jones}},
  \bibinfo{author}{\bibfnamefont{J.}~\bibnamefont{Dingjan}},
  \bibinfo{author}{\bibfnamefont{B.}~\bibnamefont{Darqui\'{e}}},
  \bibinfo{author}{\bibfnamefont{G.}~\bibnamefont{Messin}},
  \bibinfo{author}{\bibfnamefont{A.}~\bibnamefont{Browaeys}}, \bibnamefont{and}
  \bibinfo{author}{\bibfnamefont{P.}~\bibnamefont{Grangier}},
  \bibinfo{journal}{Nature} \textbf{\bibinfo{volume}{440}},
  \bibinfo{pages}{779} (\bibinfo{year}{2006}).

\bibitem[{\citenamefont{Sagioro et~al.}(2004)\citenamefont{Sagioro, Olindo,
  Monken, and P\'{a}dua}}]{sagioro_time_2004}
\bibinfo{author}{\bibfnamefont{M.~A.} \bibnamefont{Sagioro}},
  \bibinfo{author}{\bibfnamefont{C.}~\bibnamefont{Olindo}},
  \bibinfo{author}{\bibfnamefont{C.~H.} \bibnamefont{Monken}},
  \bibnamefont{and}
  \bibinfo{author}{\bibfnamefont{S.}~\bibnamefont{P\'{a}dua}},
  \bibinfo{journal}{Phys. Rev. A} \textbf{\bibinfo{volume}{69}},
  \bibinfo{pages}{053817} (\bibinfo{year}{2004}).

\bibitem[{\citenamefont{Zavatta et~al.}(2004)\citenamefont{Zavatta, Viciani,
  and Bellini}}]{zavatta_recurrent_2004}
\bibinfo{author}{\bibfnamefont{A.}~\bibnamefont{Zavatta}},
  \bibinfo{author}{\bibfnamefont{S.}~\bibnamefont{Viciani}}, \bibnamefont{and}
  \bibinfo{author}{\bibfnamefont{M.}~\bibnamefont{Bellini}},
  \bibinfo{journal}{Phys. Rev. A} \textbf{\bibinfo{volume}{70}},
  \bibinfo{pages}{023806} (\bibinfo{year}{2004}).

\bibitem[{\citenamefont{Cosme et~al.}(2008)\citenamefont{Cosme, P\'{a}dua,
  Bovino, Mazzei, Sciarrino, and De~Martini}}]{cosme_hong-ou-mandel_2008}
\bibinfo{author}{\bibfnamefont{O.}~\bibnamefont{Cosme}},
  \bibinfo{author}{\bibfnamefont{S.}~\bibnamefont{P\'{a}dua}},
  \bibinfo{author}{\bibfnamefont{F.~A.} \bibnamefont{Bovino}},
  \bibinfo{author}{\bibfnamefont{A.}~\bibnamefont{Mazzei}},
  \bibinfo{author}{\bibfnamefont{F.}~\bibnamefont{Sciarrino}},
  \bibnamefont{and}
  \bibinfo{author}{\bibfnamefont{F.}~\bibnamefont{De~Martini}},
  \bibinfo{journal}{Phys. Rev. A} \textbf{\bibinfo{volume}{77}},
  \bibinfo{pages}{053822} (\bibinfo{year}{2008}).

\bibitem[{\citenamefont{F{\`e}ve et~al.}(2007)\citenamefont{F{\`e}ve, Mah{\'e},
  Berroir, Kontos, Pla{\c c}ais, Glattli, Cavanna, Etienne, and
  Jin}}]{feve_-demand_2007}
\bibinfo{author}{\bibfnamefont{G.}~\bibnamefont{F{\`e}ve}},
  \bibinfo{author}{\bibfnamefont{A.}~\bibnamefont{Mah{\'e}}},
  \bibinfo{author}{\bibfnamefont{J.}~\bibnamefont{Berroir}},
  \bibinfo{author}{\bibfnamefont{T.}~\bibnamefont{Kontos}},
  \bibinfo{author}{\bibfnamefont{B.}~\bibnamefont{Pla{\c c}ais}},
  \bibinfo{author}{\bibfnamefont{D.~C.} \bibnamefont{Glattli}},
  \bibinfo{author}{\bibfnamefont{A.}~\bibnamefont{Cavanna}},
  \bibinfo{author}{\bibfnamefont{B.}~\bibnamefont{Etienne}}, \bibnamefont{and}
  \bibinfo{author}{\bibfnamefont{Y.}~\bibnamefont{Jin}},
  \bibinfo{journal}{Science} \textbf{\bibinfo{volume}{316}},
  \bibinfo{pages}{1169 } (\bibinfo{year}{2007}).

\bibitem[{\citenamefont{Mah{\'e} et~al.}(2010)\citenamefont{Mah{\'e},
  Parmentier, Bocquillon, Berroir, Glattli, Kontos, Pla{\c c}ais, F{\`e}ve,
  Cavanna, and Jin}}]{mahe_current_2010}
\bibinfo{author}{\bibfnamefont{A.}~\bibnamefont{Mah{\'e}}},
  \bibinfo{author}{\bibfnamefont{F.~D.} \bibnamefont{Parmentier}},
  \bibinfo{author}{\bibfnamefont{E.}~\bibnamefont{Bocquillon}},
  \bibinfo{author}{\bibfnamefont{J.}~\bibnamefont{Berroir}},
  \bibinfo{author}{\bibfnamefont{D.~C.} \bibnamefont{Glattli}},
  \bibinfo{author}{\bibfnamefont{T.}~\bibnamefont{Kontos}},
  \bibinfo{author}{\bibfnamefont{B.}~\bibnamefont{Pla{\c c}ais}},
  \bibinfo{author}{\bibfnamefont{G.}~\bibnamefont{F{\`e}ve}},
  \bibinfo{author}{\bibfnamefont{A.}~\bibnamefont{Cavanna}}, \bibnamefont{and}
  \bibinfo{author}{\bibfnamefont{Y.}~\bibnamefont{Jin}},
  \bibinfo{journal}{Phys. Rev. B} \textbf{\bibinfo{volume}{82}},
  \bibinfo{pages}{201309} (\bibinfo{year}{2010}).

\bibitem[{\citenamefont{Leicht et~al.}(2011)\citenamefont{Leicht, Mirovsky,
  Kaestner, Hohls, Kashcheyevs, Kurganova, Zeitler, Weimann, Pierz, and
  Schumacher}}]{leicht_generation_2011}
\bibinfo{author}{\bibfnamefont{C.}~\bibnamefont{Leicht}},
  \bibinfo{author}{\bibfnamefont{P.}~\bibnamefont{Mirovsky}},
  \bibinfo{author}{\bibfnamefont{B.}~\bibnamefont{Kaestner}},
  \bibinfo{author}{\bibfnamefont{F.}~\bibnamefont{Hohls}},
  \bibinfo{author}{\bibfnamefont{V.}~\bibnamefont{Kashcheyevs}},
  \bibinfo{author}{\bibfnamefont{E.~V.} \bibnamefont{Kurganova}},
  \bibinfo{author}{\bibfnamefont{U.}~\bibnamefont{Zeitler}},
  \bibinfo{author}{\bibfnamefont{T.}~\bibnamefont{Weimann}},
  \bibinfo{author}{\bibfnamefont{K.}~\bibnamefont{Pierz}}, \bibnamefont{and}
  \bibinfo{author}{\bibfnamefont{H.~W.} \bibnamefont{Schumacher}},
  \bibinfo{journal}{Semicond. Sci. Technol.} \textbf{\bibinfo{volume}{26}},
  \bibinfo{pages}{055010} (\bibinfo{year}{2011}).

\bibitem[{\citenamefont{Martin and Landauer}(1992)}]{martin_wave-packet_1992}
\bibinfo{author}{\bibfnamefont{T.}~\bibnamefont{Martin}} \bibnamefont{and}
  \bibinfo{author}{\bibfnamefont{R.}~\bibnamefont{Landauer}},
  \bibinfo{journal}{Phys. Rev. B} \textbf{\bibinfo{volume}{45}},
  \bibinfo{pages}{1742} (\bibinfo{year}{1992}).

\bibitem[{\citenamefont{Bocquillon et~al.}(2012)\citenamefont{Bocquillon,
  Parmentier, Grenier, Berroir, Degiovanni, Glattli, Pla\c{c}ais, Cavanna, Jin,
  and F\`{e}ve}}]{bocquillon_electron_2012}
\bibinfo{author}{\bibfnamefont{E.}~\bibnamefont{Bocquillon}},
  \bibinfo{author}{\bibfnamefont{F.~D.} \bibnamefont{Parmentier}},
  \bibinfo{author}{\bibfnamefont{C.}~\bibnamefont{Grenier}},
  \bibinfo{author}{\bibfnamefont{J.~M.} \bibnamefont{Berroir}},
  \bibinfo{author}{\bibfnamefont{P.}~\bibnamefont{Degiovanni}},
  \bibinfo{author}{\bibfnamefont{D.~C.} \bibnamefont{Glattli}},
  \bibinfo{author}{\bibfnamefont{B.}~\bibnamefont{Pla\c{c}ais}},
  \bibinfo{author}{\bibfnamefont{A.}~\bibnamefont{Cavanna}},
  \bibinfo{author}{\bibfnamefont{Y.}~\bibnamefont{Jin}}, \bibnamefont{and}
  \bibinfo{author}{\bibfnamefont{G.}~\bibnamefont{F\`{e}ve}},
  \bibinfo{journal}{{arXiv:1202.6243}}  (\bibinfo{year}{2012}).

\bibitem[{\citenamefont{Burkard and Loss}(2003)}]{burkard_lower_2003}
\bibinfo{author}{\bibfnamefont{G.}~\bibnamefont{Burkard}} \bibnamefont{and}
  \bibinfo{author}{\bibfnamefont{D.}~\bibnamefont{Loss}},
  \bibinfo{journal}{Phys. Rev. Lett.} \textbf{\bibinfo{volume}{91}},
  \bibinfo{pages}{087903} (\bibinfo{year}{2003}).

\bibitem[{\citenamefont{Giovannetti et~al.}(2006)\citenamefont{Giovannetti,
  Frustaglia, Taddei, and Fazio}}]{giovannetti_electronic_2006}
\bibinfo{author}{\bibfnamefont{V.}~\bibnamefont{Giovannetti}},
  \bibinfo{author}{\bibfnamefont{D.}~\bibnamefont{Frustaglia}},
  \bibinfo{author}{\bibfnamefont{F.}~\bibnamefont{Taddei}}, \bibnamefont{and}
  \bibinfo{author}{\bibfnamefont{R.}~\bibnamefont{Fazio}},
  \bibinfo{journal}{Phys. Rev. B} \textbf{\bibinfo{volume}{74}},
  \bibinfo{pages}{115315} (\bibinfo{year}{2006}).

\bibitem[{\citenamefont{F\`{e}ve et~al.}(2008)\citenamefont{F\`{e}ve,
  Degiovanni, and Jolicoeur}}]{feve_quantum_2008}
\bibinfo{author}{\bibfnamefont{G.}~\bibnamefont{F\`{e}ve}},
  \bibinfo{author}{\bibfnamefont{P.}~\bibnamefont{Degiovanni}},
  \bibnamefont{and}
  \bibinfo{author}{\bibfnamefont{T.}~\bibnamefont{Jolicoeur}},
  \bibinfo{journal}{Phys. Rev. B} \textbf{\bibinfo{volume}{77}},
  \bibinfo{pages}{035308} (\bibinfo{year}{2008}).

\bibitem[{\citenamefont{Ol{\textquoteright}khovskaya
  et~al.}(2008)\citenamefont{Ol{\textquoteright}khovskaya, Splettstoesser,
  Moskalets, and B\"{u}ttiker}}]{olkhovskaya_shot_2008}
\bibinfo{author}{\bibfnamefont{S.}~\bibnamefont{Ol{\textquoteright}khovskaya}},
  \bibinfo{author}{\bibfnamefont{J.}~\bibnamefont{Splettstoesser}},
  \bibinfo{author}{\bibfnamefont{M.}~\bibnamefont{Moskalets}},
  \bibnamefont{and}
  \bibinfo{author}{\bibfnamefont{M.}~\bibnamefont{B\"{u}ttiker}},
  \bibinfo{journal}{Phys. Rev. Lett.} \textbf{\bibinfo{volume}{101}},
  \bibinfo{pages}{166802} (\bibinfo{year}{2008}).

\bibitem[{\citenamefont{Moskalets and
  B\"{u}ttiker}(2011)}]{moskalets_spectroscopy_2011}
\bibinfo{author}{\bibfnamefont{M.}~\bibnamefont{Moskalets}} \bibnamefont{and}
  \bibinfo{author}{\bibfnamefont{M.}~\bibnamefont{B\"{u}ttiker}},
  \bibinfo{journal}{Phys. Rev. B} \textbf{\bibinfo{volume}{83}},
  \bibinfo{pages}{035316} (\bibinfo{year}{2011}).

\bibitem[{\citenamefont{Parmentier et~al.}(2011)\citenamefont{Parmentier,
  Bocquillon, Berroir, Glattli, Pla\c{c}ais, F\`{e}ve, Albert, Flindt, and
  B\"{u}ttiker}}]{parmentier_current_2011}
\bibinfo{author}{\bibfnamefont{F.~D.} \bibnamefont{Parmentier}},
  \bibinfo{author}{\bibfnamefont{E.}~\bibnamefont{Bocquillon}},
  \bibinfo{author}{\bibfnamefont{J.~M.} \bibnamefont{Berroir}},
  \bibinfo{author}{\bibfnamefont{D.~C.} \bibnamefont{Glattli}},
  \bibinfo{author}{\bibfnamefont{B.}~\bibnamefont{Pla\c{c}ais}},
  \bibinfo{author}{\bibfnamefont{G.}~\bibnamefont{F\`{e}ve}},
  \bibinfo{author}{\bibfnamefont{M.}~\bibnamefont{Albert}},
  \bibinfo{author}{\bibfnamefont{C.}~\bibnamefont{Flindt}}, \bibnamefont{and}
  \bibinfo{author}{\bibfnamefont{M.}~\bibnamefont{B\"{u}ttiker}},
  \bibinfo{journal}{{arXiv:1111.3136}}  (\bibinfo{year}{2011}).

\bibitem[{\citenamefont{Grenier et~al.}(2011)\citenamefont{Grenier, Herv\'{e},
  Bocquillon, Parmentier, Pla\c{c}ais, Berroir, F\`{e}ve, and
  Degiovanni}}]{grenier_single-electron_2011}
\bibinfo{author}{\bibfnamefont{C.}~\bibnamefont{Grenier}},
  \bibinfo{author}{\bibfnamefont{R.}~\bibnamefont{Herv\'{e}}},
  \bibinfo{author}{\bibfnamefont{E.}~\bibnamefont{Bocquillon}},
  \bibinfo{author}{\bibfnamefont{F.~D.} \bibnamefont{Parmentier}},
  \bibinfo{author}{\bibfnamefont{B.}~\bibnamefont{Pla\c{c}ais}},
  \bibinfo{author}{\bibfnamefont{J.~M.} \bibnamefont{Berroir}},
  \bibinfo{author}{\bibfnamefont{G.}~\bibnamefont{F\`{e}ve}}, \bibnamefont{and}
  \bibinfo{author}{\bibfnamefont{P.}~\bibnamefont{Degiovanni}},
  \bibinfo{journal}{New Journal of Physics} \textbf{\bibinfo{volume}{13}},
  \bibinfo{pages}{093007} (\bibinfo{year}{2011}).

\bibitem[{\citenamefont{Moskalets et~al.}(2008)\citenamefont{Moskalets,
  Samuelsson, and B\"{u}ttiker}}]{moskalets_quantized_2008}
\bibinfo{author}{\bibfnamefont{M.}~\bibnamefont{Moskalets}},
  \bibinfo{author}{\bibfnamefont{P.}~\bibnamefont{Samuelsson}},
  \bibnamefont{and}
  \bibinfo{author}{\bibfnamefont{M.}~\bibnamefont{B\"{u}ttiker}},
  \bibinfo{journal}{Phys. Rev. Lett.} \textbf{\bibinfo{volume}{100}},
  \bibinfo{pages}{086601} (\bibinfo{year}{2008}).

\bibitem[{\citenamefont{Moskalets and
  B\"{u}ttiker}(2002)}]{moskalets_floquet_2002}
\bibinfo{author}{\bibfnamefont{M.}~\bibnamefont{Moskalets}} \bibnamefont{and}
  \bibinfo{author}{\bibfnamefont{M.}~\bibnamefont{B\"{u}ttiker}},
  \bibinfo{journal}{Phys. Rev. B} \textbf{\bibinfo{volume}{66}},
  \bibinfo{pages}{205320} (\bibinfo{year}{2002}).

\bibitem[{\citenamefont{Keeling et~al.}(2008)\citenamefont{Keeling, Shytov, and
  Levitov}}]{keeling_coherent_2008}
\bibinfo{author}{\bibfnamefont{J.}~\bibnamefont{Keeling}},
  \bibinfo{author}{\bibfnamefont{A.~V.} \bibnamefont{Shytov}},
  \bibnamefont{and} \bibinfo{author}{\bibfnamefont{L.~S.}
  \bibnamefont{Levitov}}, \bibinfo{journal}{Phys. Rev. Lett.}
  \textbf{\bibinfo{volume}{101}}, \bibinfo{pages}{196404}
  (\bibinfo{year}{2008}).

\bibitem[{\citenamefont{Rychkov et~al.}(2005)\citenamefont{Rychkov, Polianski,
  and B\"{u}ttiker}}]{rychkov_photon-assisted_2005}
\bibinfo{author}{\bibfnamefont{V.~S.} \bibnamefont{Rychkov}},
  \bibinfo{author}{\bibfnamefont{M.~L.} \bibnamefont{Polianski}},
  \bibnamefont{and}
  \bibinfo{author}{\bibfnamefont{M.}~\bibnamefont{B\"{u}ttiker}},
  \bibinfo{journal}{Phys. Rev. B} \textbf{\bibinfo{volume}{72}},
  \bibinfo{pages}{155326} (\bibinfo{year}{2005}).

\end{thebibliography}

\end{document}